\documentclass[a4paper]{article}

\usepackage{subfigure}
\usepackage{graphicx}
\usepackage{authblk}
\usepackage[margin=1.in]{geometry}
\graphicspath{{images/}}
\title{Reducing audio membership inference attack accuracy to chance: 4 defenses}
\author[1]{Michael Lomnitz}
\author[1]{Nina Lopatina}
\author[2]{Paul Gamble}
\author[1]{Zigfried Hampel-Arias}
\author[1]{Lucas Tindall}
\author[1]{Felipe A. Mejia}
\author[2]{Maria Alejandra Barrios}

\affil[1]{Lab41, In-Q-Tel, Menlo Park, USA}
\affil[2]{Previously at Lab41, In-Q-Tel, Menlo Park, USA}

\begin{document}
\maketitle
\begin{abstract}
It is critical to understand the privacy and robustness vulnerabilities of machine learning models, as their implementation expands in scope. In membership inference attacks, adversaries can determine whether a particular set of data was used in training, putting the privacy of the data at risk.  
Existing work has mostly focused on image related tasks; we generalize this type of attack to speaker identification on audio samples.  We demonstrate attack precision of 85.9\% and recall of 90.8\%  for LibriSpeech, and 78.3\% precision and 90.7\% recall for VOiCES (Voices Obscured in Complex Environmental Settings). We find that implementing defenses such as prediction obfuscation, defensive distillation or adversarial training, can reduce  attack accuracy to chance. 

\end{abstract}

\section{Introduction}

The increasing adoption of machine learning (ML) algorithms requires thorough understanding of ML models’ potential vulnerabilities to attack, whether rendering a model incapable of accomplishing its task \cite{Nguyen-2016, Grosse} or extracting information from it \cite{Shokri-2016, Salem-2018, Homer, Ateniese_2015}. Deep learning (DL) models are quickly being incorporated into workflows involving audio data, in tasks such as automatic speech recognition (ASR) or speaker identification (SID). Recent research in adversarial attacks on audio data has shed light on the vulnerabilities of such systems \cite{Carlini-2018, Houdini}. Still, little is known about challenges and methods for membership inference attacks on models trained with speech data.

SID is typically implemented as a method to authenticate biometrics. Compromising the privacy of members enrolled in an SID model can not only reveal their participation in the institution using their voice as a verification key, e.g. a particular bank. It can also provide an entering point to forge and use this biometric key for nefarious purposes. Here, we discuss adversary attempts to infer the membership status of specific speakers in SID models.
Membership inference attacks were first explored in \cite{Shokri-2016}, where the authors propose using a series of attack models (one per class) to make membership inference on the output of the target model. In \cite{Salem-2018}, the authors expand on this work, showing that a successful membership inference attack can be achieved even without access to the model weights or training data.
In this paper, membership inference attacks discussed in \cite{Salem-2018} are extended to DL models in the do- main of audio data.  We study the efficacy of these types of attacks on speaker membership, on DL networks for SID. We relax the degree of the assumption of a similar data distribution since it is unrealistic that the attacker would already possess audio data from the speakers they are targeting. Here, we present attack results on two different models: one trained on the LibriSpeech \cite{Librispeech} corpus – close range microphony of a single speaker with no noise, and another using the VOiCES (Voices Obscured in Complex Environmental Settings) \cite{VOiCES} corpus – far-field recordings of a single speaker with natural reverberation and noise. This provides a point of comparison between models trained with clean, curated data, and more realistic data. We also show that we can defend against high attack accuracy.

\section{Methodology}
\subsection{Attack description}
The membership attacks implemented are based on \cite{Salem-2018}.  The adversarial attack model (A), a supervised ML binary classifier, determines whether a specific data sample was part of the target model’s (T) training data. We assume black-box access only. To gain insight into the behavior of T, the adversary trains an imitation model referred to as a shadow network (S). An adversary training S to classify examples on an independent data-set, for which training membership status is known, and having access to the posterior probabilities for in-set data for S, has all the data needed to train A. 

To simulate these attack models, we followed procedures to avoid cross- contamination between the samples used for the attacker and target, while maintaining a similar distribution. Our splits were done independently for male and female speakers, to ensure that the number of usable files are similar. We included train and validation splits for both the target and shadow network, on separate speakers. We also split data into evenly sized sets of in and out of training-set, separately for both training and evaluation the attack network.

\subsection{Model architectures}
The SID tasks are accomplished using relatively simple DL models. Three second utterances are taken from the speaker data and processed using a short-time Fourier Transformation (STFT) to extract key frequency and time features of their voices.  The STFT is used to determine the frequency content of local sections of a signal as it changes over time. In this work, we discard the relative phases and use, as input, the absolute magnitude of these spectrograms as a function of time for the models. 

Classification models: Both the target (T) and shadow (S) networks consist of three 1D convolutional blocks, followed by two fully connected layers. The target, shadow and attack models for each data-set vary only in the dimensions of the last layer, as they are trained on a slightly different number of speakers.

Attack model:  The (A) network’s input is the probability vector produced from the SID classification model, through which it can infer the membership status of a data point from the top three most likely classes. The architecture consists of two fully connected layers with 64 hidden units, with a final sigmoid layer for the binary classification. Highest overall attack accuracy is reported.

\subsection{Defenses}
To further evaluate membership inference attack threat, we tested attack network performance after implementing a series of defense mechanisms to the (T) network.
  
\emph{Adversarial  regularization}:  One reason that DL models are prone to membership inference attacks is the inherent overfitting of DL models.  A regularization  technique, discussed by \cite{Nasr-2018}, focuses not only on reducing overfitting, but explicitly training the classification network to simultaneously minimize the classification error and anticipate the membership inference attack.  This approach, modeled as a min-max privacy game, maximizes membership inference accuracy. To protect the data privacy, the gain of the inference attack is added as a regularizer for the classifier, using a regularization parameter (privacy $\lambda$) to control the trade-off between the two tasks. The two models are trained simultaneously, in a similar process as in generative adversarial networks.

\emph{Prediction obfuscation}: In a black-box scenario, inferring the membership status of examples relies heavily on the probability distribution of the predictions made by the target model (T), where particularly confident predictions indicate that T has likely seen the data point during training. Prediction obfuscation defenses block this approach, defending the privacy of the data by reducing the amount of information provided to the attacker. We study two facets of this defense: reducing the number predictions provided to the attacker (i.e. probability of the top-k classes instead of the entire vector), and returning the classes by rank, instead of their actual probabilities.

\emph{Distillation training}: Though model distillation \cite{Hinton-2015} as a defense \cite{Papernot-2015} was initially proposed to protect against model fooling via adversarial perturbations, we explore it's use against membership inference attacks. In distillation training, a DL network is trained using knowledge transferred from a different network.  This defense is motivated by how knowledge acquired by a network is not only encoded in the weights in the model, but is also reflected in the network’s probability vectors. The procedure allows the model to learn using additional information about the classes, encouraging more effective generalization. This forces the network to produce probability vectors with large values for each class, breaking the ability of the attack network to use this information.

\emph{Model key}: Another strategy to defend the privacy of data is to allow the model to return erroneous results when we expect it is being targeted. As an example we explore a specific approach: random noise is sampled from a uniform distribution in the range [0,1] and randomly added to the data during training. The SID system is trained to identify the speaker in the presence of this noise and a second model is trained to identify whether or not there is noise present. During inference, for samples where the noise has not been added the model returns a random prediction instead of the original output of the model.

\section{Results}
\subsection{Target and Attack baseline}
Tables \ref{tab:libri_baselines} and \ref{tab:voices_baselines} show the baseline results of our studies on  LibriSpeech and VOiCES data-sets for our baseline models ($\alpha = 0$) as well as with different degrees of $L2$ regularization.  In networks trained on the LibriSpeech data-set, we see a small reduction in the attack performance as the value of $\alpha$ is increased, with no significant change in the target accuracy. In networks trained on the VOiCES data-set, we see a larger reduction in attack performance, as well as a notable reduction in the target model accuracy.

To test the validity of our SID model, we also performed enrollment in one room and verification in a second room. Our results for enrollment in room 1 and verification in room 2 were 98.43\% training and 58.43\% test accuracy. Enrollment in room 2 and verification in room 1 results were 96.33\% training and 75.87\% test accuracy. Our results show considerable variance between the  verification but not the enrollment between these two rooms. However, our results for enrollment in room 2 and verification in room 1 are very comparable to the room 1 test \& train. 

\begin{table}[th]
  \caption{Summary of target and attack baseline results for LibriSpeech SID models.}
  \label{tab:libri_baselines}
  \centering
  \begin{tabular}{ c c c c c c }
    \hline
     & \multicolumn{2}{ c }{\centering Target acc.} & \multicolumn{3}{ c }{\centering Attack performance} \\ 
    \hline
    Data & Train  & Test  &  Acc. & Prec. & Rec. \\ 
     \hline
    $\alpha$ & Train  & Test  &  Acc. & Prec. & Rec. \\ 
    0 & 99.5 & 95.0 & 87.9 & 85.8 & 90.8 \\
    0.001 & 99.4 & 94.6 & 89.4 & 89.7 & 89.5 \\
    0.005 & 99.4 & 95.2 & 87.3 & 97.4 & 77.1 \\
    0.01 & 99.0 & 93.8 & 81.6 & 98.3 & 64.9 \\ 
    \hline
  \end{tabular}
  
\end{table}

\begin{table}[th]
  \caption{Summary of target and attack baseline results for VOiCES SID models.}
  \label{tab:voices_baselines}
  \centering
  \begin{tabular}{ c c c c c c }
    \hline
     & \multicolumn{2}{ c }{\centering Target acc.} & \multicolumn{3}{ c }{\centering Attack performance} \\ 
    \hline
    Data & Train  & Test  &  Acc. & Prec. & Rec. \\ 
     \hline
    $\alpha$ & Train  & Test  &  Acc. & Prec. & Rec. \\ 
    0 & 98.6 & 77.8 & 82.8 & 78.3 & 90.7 \\ 
    0.001 & 88.4 & 72.9 & 66.9 & 93.0 & 36.6 \\
    0.005 & 83.0 & 69.1 & 55.8 & 93.9 & 12.5 \\
    0.01 & 72.4 & 61.1 & 51.2 &  95.6 & 3.66 \\ 
    \hline
  \end{tabular}
  
\end{table}

\subsection{Adversarial Training}
Table \ref{tab:adversarial_training} summarizes results obtained on the LibriSpeech data-set through adversarial regularization for different values of the privacy parameter, $\lambda$. Because of time/computing constraints, we only tested this defense on the smaller, LibriSpeech data-set. Including the loss from the inference attack dramatically slows the overall training of the target model, with a minor effect on overall performance. We see relatively small changes for the potency of the inference attack for small values of the privacy parameter λ.  However, the model trained with $\lambda$ = 3 is considerably more robust to membership inference attack. After adversarial training, the attack accuracy drops to slightly better than random (~60\%) with only minor effects on its overall performance.  

\begin{table}[th]
  \caption{Attack performance following adversarial training for different values of the privacy constant $\lambda$ on LibriSpeech SID model.}
  \label{tab:adversarial_training}
  \centering
  \begin{tabular}{ c c c c c c }
    \hline
     & \multicolumn{2}{ c }{\centering Target acc.} & \multicolumn{3}{ c }{\centering Attack performance} \\ 
    \hline
     & Train  & Test  &  Acc. & Prec. & Rec. \\ 
     \hline
    $\lambda$ & Train  & Test  &  Acc. & Prec. & Rec. \\ 
    1.0 & 96.3 & 89.0 & 81.2 & 87.5 & 73.6 \\ 
    2.0 & 95.9 & 87.2 & 80.6 & 81.0 & 80.7 \\ 
    3.0 & 91.1 & 80.4 & 63.4 & 93.6 & 30.0 \\ 
    \hline
  \end{tabular}
  
\end{table}

\subsection{Prediction obfuscation}
In table \ref{tab:obfuscation}, we show the effect of applying prediction obfuscation on the outputs of target models. Given that the attack relies heavily on the probability vectors to infer membership status, replacing probability vectors with a simple ranking nullifies the attack without affecting the model accuracy. However, this limits the amount of information provided to good actors, which could potentially have negative effects in real-life scenarios.

\begin{table}[th]
  \caption{Attack performance following adversarial training for different values of the privacy constant $\lambda$ on LibriSpeech SID model.}
  \label{tab:obfuscation}
  \centering
  \begin{tabular}{ c c c c c c }
    \hline
     & \multicolumn{2}{ c }{\centering Target acc.} & \multicolumn{3}{ c }{\centering Attack performance} \\
    \hline
    Data & Train  & Test  &  Acc. & Prec. & Rec. \\  
     \hline
    Libri & 99.5 & 95.0 & 49.8 & 0 & 0 \\ 
    VOiCES & 98.6 & 77.8 & 50.0 & 0 & 0 \\ 
    \hline
  \end{tabular}
  
\end{table}

\subsection{Defensive distillation}
Tables \ref{tab:distillation_libri} and \ref{tab:distillation_voices} show the attack performance on target models trained with defensive distillation at different distillation temperatures $T$ for LibriSpeech and VOiCES data. With increasing temperature, the LibriSpeech model obtained greater privacy in exchange with only a minor performance trade-off.

\begin{table}[th]
  \caption{Attack performance performance following defensive distillation  at different values of the distillation temperature $T$ on LibriSpeech SID model.}
  \label{tab:distillation_libri}
  \centering
  \begin{tabular}{ c c c c c c }
    \hline
     & \multicolumn{2}{ c }{\centering Target acc.} & \multicolumn{3}{ c }{\centering Attack performance} \\
    \hline
    $T$ & Train  & Test  &  Acc. & Prec. & Rec. \\ 
     \hline
    1 & 97.6 & 92.0 & 83.9 & 87.8 & 79.4 \\ 
    5 & 97.5 & 91.0 & 75.8 & 69.6 & 93.1 \\ 
    10 & 96.9 & 91.3 & 67.9 & 62.1 & 94.9 \\ 
    100 & 96.8 & 89.1 & 52.8 & 51.9 & 99.6 \\ 
    \hline
  \end{tabular}
  
\end{table}

\begin{table}[th]
  \caption{Attack performance performance following defensive distillation  at different values of the distillation temperature $T$ on VOiCES SID model.}
  \label{tab:distillation_voices}
  \centering
  \begin{tabular}{ c c c c c c }
    \hline
     & \multicolumn{2}{ c }{\centering Target acc.} & \multicolumn{3}{ c }{\centering Attack performance} \\
    \hline
    $T$ & Train  & Test  &  Acc. & Prec. & Rec. \\ 
     \hline

    1.0 & 95.6 & 80.3 & 80.6 & 83.1 & 76.7 \\ 
    5.0 & 96.6 & 80.9 & 73.8 & 67.4 & 92.1 \\ 
    10.0 & 95.4 & 79.4 & 63.8 & 58.5 & 95.3 \\
    100.0 & 95.6 & 81.2 & 51.3 & 50.7 & 99.5 \\ 
    \hline
  \end{tabular}
  
\end{table}

A similar pattern holds for the defenses for models trained on VOiCES. However, in contrast to the LibriSpeech model, the VOiCES SID model improved in accuracy ($\sim$5\%) at all $T$. It is interesting that the distillation procedure encourages the target to generalize better than the baseline model.

\subsection{Model key}
Table \ref{tab:key} summarizes the attack performance on a model expecting inputs from a distribution shifted slightly by the addition of random noise. In both cases, the model accuracy in training and validation drops (1-3 \% LibriSpeech \& ~9 \% VOiCES), however as expected the membership inference attack is nullified and the accuracy drops to random.  

Though effective, this defense restricts the number of users who could use the model as they would require knowledge on how to pre-process the data to obtain results of significance.  This is illustrated by the model train and validation accuracy without the required noise( quantities in brackets in table \ref{tab:key}).

\begin{table}[th]
  \caption{Attack performance on both LibriSpeech and VOiCES on the model using model key. The values in brackets indicate the target models performance on data without noise.}
  \label{tab:key}
  \centering
  \begin{tabular}{ c c c c c c }
    \hline
     & \multicolumn{2}{ c }{\centering Target acc.} & \multicolumn{3}{ c }{\centering Attack performance} \\
    \hline
    $\lambda$ & Train  & Test  &  Acc. & Prec. & Rec. \\  
     \hline
    Libri & 98.0 (0.8) & 92.3 (0.8) & 50.3 & 50.2 & 83.7 \\ 
    VOiCES & 93.4 (0.9) & 68.5 (0.8) & 50.0 & 50.0 & 88.4 \\ 
    \hline
  \end{tabular}
\end{table}

\section{Conclusions}

\begin{figure*}[hb!]

\begin{center}
\centerline{
    \includegraphics[width=0.7\textwidth]{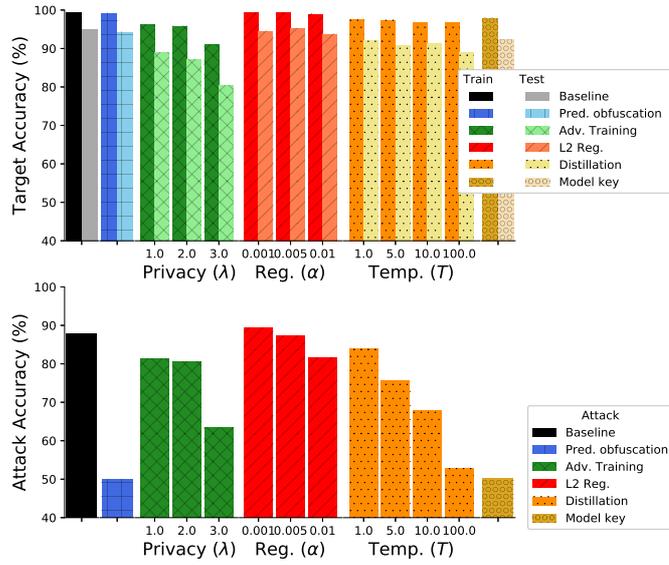}
}
\caption{Summary of adversary 1 results obtained on LibriSpeech for the baseline and defended models. The top bar plot shows the target train (dark) and test (light) accuracy. The bottom plot shows the attack performance.}
\label{fig:libri_results}
\end{center}

\end{figure*}

\begin{figure*}[hb!]

\begin{center}
\centerline{
    \includegraphics[width=0.7\textwidth]{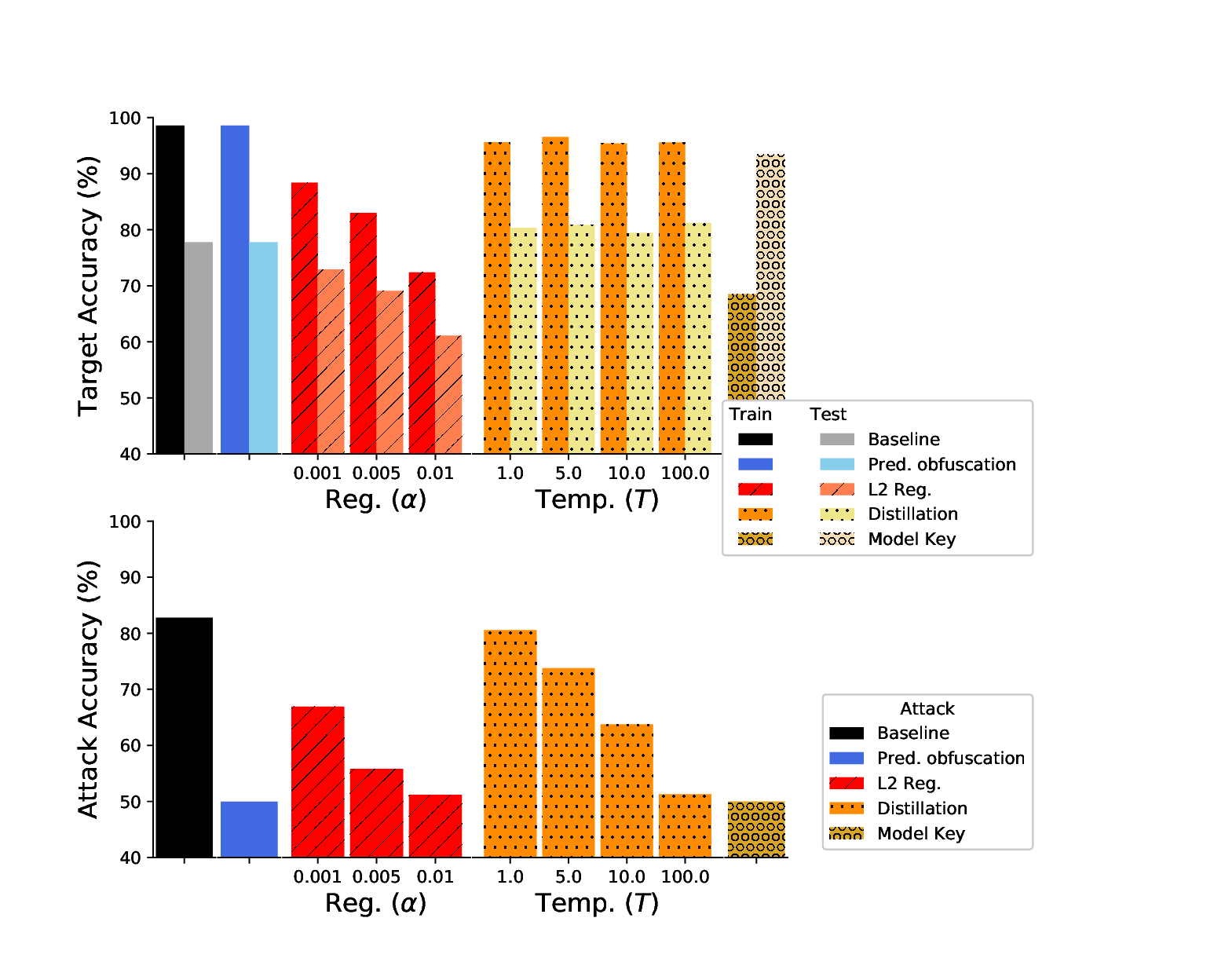}
}
\caption{Summary of adversary 1 results obtained on VOiCES for the baseline and defended models. The top bar plot shows the target train (dark) and test (light) accuracy. The bottom plot shows the attack performance.}
\label{fig:voices_results}
\end{center}

\end{figure*}

In this report, we delineated, implemented, and assessed the first membership inference attack against speech data. Figures \ref{fig:libri_results} \& \ref{fig:voices_results} compare all of our results as discussed in the previous section. All three of our defenses reduced attack accuracy near chance, two lowering nearly exactly to chance. The most effective defenses for LibriSpeech, with the lowest effect on target performance, were defensive distillation and predictive obfuscation, allowing users to choose a disadvantage: a slight reduction in accuracy or reduced granularity of results, respectively. Interestingly, the optimal defense for VOiCES, defensive distillation, both reduced attack accuracy to chance, and increased the target model's accuracy. This suggests that membership inference defenses may be more effectively deployed on models trained with realistic, noisy audio data. Since greater background noise adds variance to the data, methods that thwart membership inference by increasing generalization may not require a trade-off between safety and performance. 
We showed the high vulnerability of SID models to membership inference attack: 87.9 \% accuracy for LibriSpeech and 82.8 \% for VOiCES. Membership in speech data-sets is sensitive in any system which uses voice as a biometric marker, or a system whose training data must otherwise be kept private. Therefore, these results have significant practical privacy implications.
Reassuringly, we next showed that these attacks can be defended with minimal to no performance decrement of the target model. The results for LibriSpeech and VOiCES trained models further suggest that defenses must be tested for each model and threat type individually. It is therefore critical to further expand the scope of research into privacy attacks to additional domains (e.g. facial recognition, text, biological data), \& more realistic data. 

\clearpage
\bibliographystyle{IEEEtran}

\bibliography{mybib}

\end{document}